\documentclass[12pt]{iopart}

\usepackage{iopams}  
\usepackage{setstack}  
\usepackage{graphicx,color}  

\newtheorem{example}{Example}

\begin{document}


\title{Statistics and Nos\'e formalism for Ehrenfest dynamics}

\author{J L Alonso,$^{1,2,3}$ A Castro,$^2$ J Clemente-Gallardo,$^{1,2,3}$ J C Cuch\'{\i},$^4$ P Echenique$^{5,1,2,3}$ and F Falceto$^{1,2}$}
\address{$^1$Departamento de F\'{\i}sica Te\'orica, Universidad de Zaragoza, 
  Campus San Francisco, 50009 Zaragoza (Spain)}
\address{$^2$Instituto de Biocomputaci{\'{o}}n y F{\'{\i}}sica de Sistemas Complejos (BIFI), 
  Universidad de Zaragoza,  Edificio I+D, Mariano Esquillor s/n, 50018
  Zaragoza (Spain)}
\address{$^3$Unidad asociada IQFR-BIFI, Spain}
\address{$^4$Departament d'Enginyeria Agroforestal, ETSEA- Universitat de Lleida,
  Av. Alcalde Rovira Roure 191, 25198 Lleida  (Spain)}
\address{$^5$Instituto de Qu\'{\i}mica F\'{\i}sica ``Rocasolano''
  (CSIC), C/ Serrano 119, 28006 Madrid (Spain)}
\ead{alonso.buj@gmail.com}
\ead{acastro@bifi.es}
\ead{jesus.clementegallardo@bifi.es}
\ead{cuchi@eagrof.udl.cat}
\ead{echenique.p@gmail.com}
\ead{falceto@unizar.es}

\begin{abstract}
  Quantum dynamics (i.e., the Schr{\"{o}}dinger equation) and
  classical dynamics (i.e., Hamilton equations) can both be formulated
  in equal geometric terms: a Poisson bracket defined on a manifold.
  In this paper we first show that the hybrid quantum-classical
  dynamics prescribed by the Ehrenfest equations can also be
  formulated within this general framework, what has been used in the
  literature to construct propagation schemes for Ehrenfest dynamics.
  Then, the existence of a well defined Poisson bracket allows to
  arrive to a Liouville equation for a statistical ensemble of
  Ehrenfest systems.  The study of a generic toy model shows that the
  evolution produced by Ehrenfest dynamics is ergodic and therefore
  the only constants of motion are functions of the Hamiltonian. The
  emergence of the canonical ensemble characterized by the Boltzmann
  distribution follows after an appropriate application of the
  principle of equal a priori probabilities to this case.  Once we
  know the canonical distribution of a Ehrenfest system, it is
  straightforward to extend the formalism of Nos{\'{e}} (invented to
  do constant temperature Molecular Dynamics by a non-stochastic
  method) to our Ehrenfest formalism. This work also provides the
  basis for extending stochastic methods to Ehrenfest dynamics.
\end{abstract}


\pacs{31.15.xv, 31.15.xr,02.70.Ns}
\submitto{\JPA}
\maketitle

\section{Introduction}
\label{sec:introduction}

The Schr{\"{o}}dinger equation for a combined system of electrons
and nuclei enables us to predict most of the chemistry and molecular
physics that surrounds us, including bio-physical processes of great
complexity. Unfortunately, this task is not possible in general, and
approximations need to be made; one of the most important and
successful being the classical approximation for a number of the
particles. Mixed quantum-classical dynamical (MQCD) models are
therefore necessary and widely used.

The so-called ``Ehrenfest equations'' (EE) result from a
straightforward application of the classical limit to a portion of the
particles of a full quantum system, and constitute the most evident
MQCD model, as well as a first step in the intricated problem of
mixing quantum and classical dynamics. For example, it can be noted
that much of the field of molecular dynamics (MD) --whether ab initio
or not-- is based on equations that can be obtained from further
approximations to the Ehrenfest model: for instance, written in the
adiabatic basis, the Ehrenfest dynamics collapse into Born-Oppenheimer
MD if we assume the non-adiabatic couplings to be negligible.

It is not the purpose of this work, however, to dwell into the
justification or validity conditions of the EE (see for example
Refs.~\cite{marxhutter:2009, gerber1982time, gerber1988self,
  bornemann1996quantum, bornemann1995pre} for rigurous analyses). Nor
is it to investigate the unsettled problem of which is the best manner
of mixing quantum and classical degrees of freedom. The Ehrenfest
model has a proven application niche, and for this reason we are
interested in investigating some of its theoretical foundations, in a
manner and with an aim that we describe in the following. For recent 
progress in non-adiabatic electronic dynamics in MQCD see, for example, 
Ref.~\cite{Zhu2005JCTC}

Classical mechanics (CM) can be formulated in several mathematical
frameworks, each corresponding to a different level of abstraction
(Newton equations, the Hamiltonian formalism, the Poisson brackets,
etc.). Perhaps its more abstract and general formulation is
geometrical, in terms of Poisson manifolds. Similarly, quantum
mechanics (QM) can be formulated in different ways, some of which
resemble its classical counterpart. For example, the observables
(self-adjoint linear operators) are endowed with a Poisson algebra
almost equal to the one that characterizes the dynamical variables in
CM.  Moreover, Schr{\"{o}}dinger equation can be recast into
Hamiltonian equations form~ (see \cite{heslot1985quantum}) by transforming
the complex Hilbert space into a real one of double dimension; the
observables are also transformed into dynamical functions in this new
phase space, in analogy with the classical case. Finally, a Poisson
bracket formulation has also been established for QM, which permits to
classify both the classical and the quantum dynamics under the same
heading.

This variety of formulations does not emerge from academic caprice;
the successive abstractions simplify further developments of the
theory, such as the step from microscopic dynamics to statistical
dynamics: the derivation of Liouville equation (or von Neumann
equation in the quantum case), at the heart of statistical dynamics,
is based on the properties of the Poisson algebra.

The issue regarding what is the correct equilibrium distribution of a
mixed quantum-classical system is seen as a very relevant one
~\cite{Par2006JCTC, Kab2006JPCA, Bas2006CPL, Kab2002PRE, Tul1998Book,
  Mul1997JCP, Mau1993EPL, Ter1991JCP,Alonso:2010p6480}. An attempt to its derivation
can be found in Ref.~\cite{Mau1993EPL}, where they arrive to the same
distribution that we will advocate below, although it is found using
the Nos{\'{e}}-Hoover technique \cite{nose:1984,nose:1991}, which in
principle is only a mathematical scheme to produce the equilibrium,
and which the very authors of Ref.~\cite{Mau1993EPL} agree that it is
not clear how to apply to a mixed quantum-classical dynamics. On the
other hand, in Ref.~\cite{Par2006JCTC} they provide some analytical
results about this distribution, but not in the case in which the
system of interest is described by the EE; instead (and as in
Refs.~\cite{Kab2006JPCA, Bas2006CPL}), they assume that the system is
fully quantum, and that it is coupled to an infinite classical bath
via an Ehrenfest-like interaction.

It is therefore necessary to base Ehrenfest dynamics --or any other
MQCD model-- on firm theoretical ground. In particular, we are
interested in establishing a clear path to statistical mechanics for
Ehrenfest systems, which in our opinion should be done by first
embedding this dynamics into the same theoretical framework used in
the pure classical or quantum cases (i.e., Poisson brackets,
symplectic forms, etc). Then the study of their statistics will follow
the usual steps for purely classical or purely quantum ensembles. In
this respect, it should be noted that other approaches to MQCD (not
based on Ehrenfest equations) exist \cite{Kisil:fk, prezhdo1997mixing,
  kapral1999mixed}, and to their corresponding statistics
~\cite{nielsen2001statistical,Kapral:2001fk}, however it has been
found that the formulation of well defined quantum-classical brackets
(i.e., satisfying the Jacobi identity and Leibniz derivation rule) is
a difficult issue
~\cite{kisil2005quantum,agostini2007we,kisil2010comment,agostini2010reply}.
On the contrary, as shown below, and as a result of the fact that the
evolution of a quantum system can be formulated in terms of Hamilton
equations similar to those of a classical system
\cite{bornemann1996quantum,Schmitt1996JCP}, we will not have
difficulties in a rigorous formulation of the Ehrenfest dynamics in
terms of Poisson brackets.

The roadmap of this project is the following: In
Section~\ref{sec:ehrenfest} we recall the definition of the Ehrenfest
model. In Section \ref{sec:append-geom-dynam-PB} we quickly summarize
the formulation of CM in terms of Poisson brackets. Then we summarize,
in Section~\ref{sec:append-geom-quant}, the analogous description of
QM in terms of geometrical objects which can be found, for example, in
Refs.~\cite{
  kibble1979geometrization,heslot1985quantum,abbati1984pure,
  cirelli1991quantum,brody2001geometric,Ashtekar:1998p906,
  Carinena:2006p7565,Carinena:2007p813,clemente2008basics}. These
works demonstrate how Schr{\"{o}}dinger equation can be written as a
set of (apparently classical) Hamiltonian equations. Also, by using a
suitable definition of observables as functions on the (real) set of
physical states, Schr\"odinger equation can be written in terms of a
canonical Poisson bracket. QM appears in this way as ``classical'',
although the existence of extra algebraic structure encodes the
probabilistic interpretation of measurements, and the superposition
and indetermination principles \cite{
  kibble1979geometrization,abbati1984pure,
  cirelli1991quantum,brody2001geometric,Ashtekar:1998p906}. 
As we have in mind the application to computer simulations, we will
consider only finite dimensional quantum systems, obtained by a
suitable sample of the electronic states. This simplifies
significantly  the description, even if the framework is valid also
for the infinite dimensional case (see \cite{Ashtekar:1998p906}).

 For the
case of Ehrenfest dynamics, we can now combine it
(Section~\ref{sec:our-proposal}) with CM, whose Poisson bracket
formulation is very well known.  This is the first of the
contributions of the paper: once we realize that the geometric
description of classical and quantum dynamics are formally analogous,
we can combine them as we do when combining different classical
systems. The procedure is very simple: we 
consider as a global phase space the Cartesian product of the
electronic and the nuclear phase spaces and define a global Poisson
bracket simply as the sum of the classical and the quantum ones. It is
completely straightforward to prove that such a Poisson bracket is
well defined and that it provides the correct dynamical equations
(EE). From a formal point of view, the resulting dynamics is more similar to
a classical system than to a quantum one, although when considering a
pure quantum system the dynamics is the usual Schr\"odinger one. This
is not surprising since the coupling of the classical and quantum
systems makes the total system nonlinear in its evolution, and this is
one of the most remarkable differences between a classical and a
quantum system in any formulation.

Once the dynamical description as a Poisson system is at our disposal,
it is a very simple task (Section~\ref{sec:defin-stat-syst}) to
construct the corresponding statistical description following the
lines of Refs.~\cite{balescu1997statistical} and
\cite{balescu1975equilibrium}.    This is another
main result of the paper: not having proved the Hamiltonian nature of
Ehrenfest dynamics, it was not possible to provide a self-consistent
definition of the Statistics corresponding to it. Once the Hamiltonian
description is available, the definition of this Statistics is straightforward.

 The formal similarity with the CM
case ensures the correctness of the procedure and allows us to derive
a Liouville equation.  The choice of the equilibrium distribution is
based, as usual, in the principle of equal \textit{a priori}
probablities implicitely used by Gibbs and clearly formulated by
Tolman \cite{tolman}. After this, in Section \ref{sec:application:-nos-e}
we extend Nos\'e  formalism to the Ehrenfest MQCD framework and
provide then a method to simulate the canonical ensemble of Ehrenfest
systems. This is the final main contribution of the paper.

Please note that for systems where the Born-Oppenheimer approximation is not
sufficient, the use of the canonical ensemble of a mixed quantum-classical system such
as Ehrenfest's is compulsory, since those systems where the electronic gap is small can
be affected by electronic temperature effects of the same order. Therefore, a simple
mechanism of simulation such as Nos\'e is very important for practical applications.

Finally, in the conclusions (Section \ref{sec:concl-future-work}), we
propose to extend, within our formalism, stochastic methods to
Ehrenfest dynamics.

\section{The Ehrenfest model}
\label{sec:ehrenfest}

The Ehrenfest Equations have the following general form:
\begin{eqnarray}
\dot{\vec{R}}_J(t) & = & \langle \psi(t) \vert \frac{\partial
  \hat{H}}{\partial \vec P_J}(\vec R(t),\vec P(t),t) \vert \psi(t)\rangle,
\\
\dot{\vec{P}}_J(t) & = & -\langle \psi(t) \vert \frac{\partial
  \hat{H}}{\partial \vec R_J}(\vec R(t),\vec P(t),t) \vert \psi(t)\rangle,
\\
i\hbar \frac{{\rm d}}{{\rm d}t} |\psi(t)\rangle & = & \hat{H}(\vec R(t), \vec
P(t),t)\vert\psi(t)\rangle,
\label{eq:ehrenfest}
\end{eqnarray}
where $(\vec R, \vec P)$ denote collectively the set of canonical
position and momenta coordinates of a set of classical particles,
whereas $\psi$ is the wavefunction of the quantum part of the system
(see Ref ~\cite{Andrade:2009p7108} for the issue of the
Hellmann-Feynman theorem in this context, i.e., whether one should
take the derivative inside or outside the expectation value).

This and other MQCD models appear in different contexts; in many
situations, the division into quantum and classical particles is made
after the electrons have been integrated out, and is used to
\emph{quantize} a few of the nuclear degrees of freedom. However, a
very obvious case to use EE is when we want to treat electrons quantum
mechanically, and nuclei classically. In order to fix ideas, let us
use this case as an example: the Hamiltonian for the full quantum
system is:
\begin{eqnarray}
  \label{eq:29a}
  \hat H &:=-\hbar^2\sum_J\frac 1{2M_J}\nabla_J^2-\hbar^2\sum_j\frac 12
  \nabla_j^2+\frac 1{4\pi \epsilon_0}\sum_{J<K}\frac{Z_JZ_K}{|\vec R_J-  
  \vec R_K|}
  \nonumber
  \\
  &\quad\;\,-\frac 1{4\pi \epsilon_0}\sum_{j<k}\frac 1{|\vec r_j-\vec r_k|}-\frac
1{4\pi \epsilon_0}\sum_{J,j}\frac {Z_J}{|\vec R_J-\vec r_j|} 
  \nonumber
  \\
  &=:-\hbar^2\sum_J\frac 1{2M_J}\nabla_J^2 -\hbar ^2\sum_j\frac 12
 \nabla_j^2+V_{n-e}(\vec r,\vec R) 
 \nonumber
 \\
 &=:-\hbar^2\sum_J\frac 1{2M_J}\nabla_J^2+H_e(\vec r, \vec R),
\end{eqnarray}
where all sums must be understood as running over the whole natural
set for each index. $M_J$ is the mass of the J-th nucleus in units of
the electron mass, and $Z_J$ is the charge of the J-th nucleus in
units of (minus) the electron charge. Also note that we have defined
the nuclei-electrons potential $V_{n-e}(\vec r ,\vec R)$ and the
electronic Hamiltonian $H_e(\vec r, \vec R)$ operators.

The EE may then be reached in the following way
\cite{marxhutter:2009,gerber1982time,gerber1988self,bornemann1996quantum,bornemann1995pre}: 
first, the full wave function is split into a product of nuclear and
electronic wave functions, which leads to the time dependent
self-consistent field model, in which the two subsystems are quantum
and 
coupled. Afterwards, a classical limit procedure is applied to the
nuclear subsystem, and the EE emerge naturally.
In terms of the nuclei positions $\vec R_{J}$ and of the element of the Hilbert space 
$|\psi\rangle \in \mathcal{H}$ which encodes the state of the
electrons of the system, the Ehrenfest dynamics is then given by
\begin{eqnarray}
  \label{eq:21a}
  M_{J}\ddot{\vec R}_{J}&=-\langle\psi | \nabla_{J}H_e(\vec r, \vec R) |\psi \rangle,
 \\
  i\hbar\frac{d}{dt} |\psi\rangle&=H_e(\vec r, \vec R) |\psi\rangle.
\end{eqnarray}

These equations can be given a Hamiltonian-type description by introducing a Hamiltonian function of the form: 
\begin{equation}
  \label{eq:22a}
  H(\vec R, \vec P)=\sum_{J}\frac{\vec P_{J}^2}{2M_J}+\langle\psi |
  H_e(\vec r,\vec R) |\psi \rangle. 
\end{equation}

Then, by fixing a relation of the form $ \vec P_{J}= M \dot {\vec
  R}_{J}$, we obtain a structure similar to Hamilton equations:
\begin{eqnarray}
  \label{eq:25a}
  \dot{\vec R}_{J}&=\frac {\vec P_{J}}{M_{J}},
  \\
  \dot {\vec P}_{J}&=-\langle\psi | \nabla_{J}H_e(\vec r, \vec R) |\psi \rangle,
  \\
 i\hbar \frac{d}{dt} |\psi\rangle&=H_e(\vec r, \vec R) |\psi\rangle.
\end{eqnarray}

This set of equations perhaps constitutes the most evident MQCD model, as well as a
first step in the intricated problem of mixing quantum and classical
dynamics. In this sense, it is interesting to note that written in the
adiabatic basis $\lbrace\psi_m(\vec{R})\rbrace_m$ formed by the
eigenvectors of the electronic Hamiltonian,
\begin{equation}
\hat{H}_{\rm e}(\vec{r}, \vec{R})\vert \psi_m(\vec{R})\rangle = 
E_m(\vec{R})\vert\psi_m(\vec{R})\rangle\,,
\end{equation}
the Ehrenfest dynamics collapse into ground state Born-Oppenheimer MD
(gsBOMD) if we assume the non-adiabatic couplings to be
negligible. Indeed, by expanding the wave function in this basis (some
relevant considerations related to this change of variables can be
found in Ref.~\cite{Andrade:2009p7108}, Section 2 ``Ehrenfest Dynamics:
  Fundaments and implications for First Principles Simulations''),
\begin{equation}
\vert\psi(t)\rangle = \sum_m c_m(t)\vert\psi_m(\vec{R}(t))\rangle\,.
\end{equation}
the equations are transformed into:
\begin{eqnarray}
\nonumber
M_J \ddot{\vec{R}}_J(t) & = & 
-\sum_m \vert c_m(t)\vert^2 \nabla_J E_m(\vec{R}(t))
\\
& & 
- \sum_{mn} c^*_m(t)c_n(t) \left[ E_m(\vec{R}(t)) - E_n(\vec{R}(t)) \right] \vec{d}_J^{mn}(\vec{R}(t))\,.
\\
i \frac{{\rm d}}{{\rm d}t} c_m(t) & = & E_m(\vec{R}(t)) c_m(t) 
-i \sum_n c_n(t) \left[
\sum_J \dot{\vec{R}}_J \cdot \vec{d}_J^{mn}(\vec{R}(t))
\right]
\end{eqnarray}
where the ``non-adiabatic couplings'' are defined as:
\begin{equation}
\vec{d}_J^{mn}(\vec{R}) :=  \langle \psi_m(\vec{R})
 \vert \nabla_J \psi_n(\vec{R}) \rangle\,.
\end{equation}
If these are negligible, and we assume that the electronic system starts
from the ground state ($c_m(0)=\delta_{m0}$), the EE model reduces to gsBOMD:
\begin{eqnarray}
M_J \ddot{\vec{R}}_J(t) & = & 
\nabla_J E_0(\vec{R}(t))\,,
\\
c_m(t) & = & \delta_{m0}\,.
\end{eqnarray}


In spite of the formal similarities, the EE equations do not correspond
yet to Hamilton equations, since they lack a global phase space
formulation (encompassing both the nuclear and the electronic degrees
of freedom) and a Poisson bracket.

\section{Classical mechanics in terms of Poisson brackets}
\label{sec:append-geom-dynam-PB}

We begin by recalling very quickly the Hamiltonian formulation of
classical dynamics (we address the interested reader to a classical
text, such as Ref.~\cite{Abraham:1978p1131}, for a detailed
presentation). Let us consider a classical system with phase space
$M_C$, which, for the sake of simplicity, we can identify with
$\mathbb{R}^{2n}$, where $n$ is the number of degrees of freedom
(strictly speaking, $M_C$ is a general $2n$-dimensional manifold,
homeomorphic to $\mathbb{R}^{2n}$ only locally). The $2n$ dimensions
correspond with the $n$ position coordinates that specify the
configuration of the system, and the $n$ corresponding momenta
(mathematically, however, the division into ``position'' and
``momenta'' coordinates is a consequence of Darboux theorem -- see
below).

The ``observables'' in classical mechanics are differentiable functions
\begin{equation}
  \label{eq:37}
  f:M_C\to \mathbb{R},
\end{equation}
that assign the result of a measurement to every point in $M_C$.  On
this set of functions $C^\infty(M_C)$ we introduce the 
\textbf{Poisson bracket}, $\{\cdot,\cdot\}$, a bilinear operation 
\begin{equation}
  \label{eq:36}
  \{ \cdot, \cdot\}:C^\infty(M_C)\times C^\infty(M_C)\to C^\infty(M_C), 
\end{equation}
which:
\begin{itemize}
\item is antisymmetric, 
$$
\{f, g\}=-\{g, f\}, \qquad \forall f,g\in C^\infty(M_C)\,,
$$
\item satisfies the Jacobi identity, i.e., $\forall
f,g,h\in C^\infty(M_C)$:
$$
\{ f, \{g, h\}\}+ \{ h, \{f, g\}\}+\{ g, \{h, f\}\}=0\,,
$$
\item and satisfies the Leibniz rule ,i.e., $\forall
f,g,h\in C^\infty(M_C)$:
$$
\{f, gh\}=\{f,g\}h+g\{f,h\}\,.
$$
\end{itemize}
If the Poisson bracket is non degenerate (it has no Casimir
functions), a theorem due to Darboux ensures that there exists a set
of coordinates $(\vec{R},\vec{P})$ for which the bracket has the ``standard'' form
(at least locally, in a neighbourhood of every point):
\begin{equation}
\{ f_1,f_2\}=\sum_{k=1}^n \frac{\partial f_1}{\partial P_k}\frac{\partial
    f_2}{\partial R^k}- \frac{\partial f_1}{\partial R^k}\frac{\partial
    f_2}{\partial P_k}.
\end{equation}
These coordinates are called Darboux coordinates. They are specially
useful when studying the dynamics and invariant measures of a
Hamiltonian system \cite{Abraham:1978p1131}.  In the rest of the paper
we shall be working with this kind of coordinates.

The Poisson bracket allows us to introduce the concept of
Hamiltonian vector field: Given a function $f\in C^\infty(M_C)$ and a Poisson bracket
  $\{\cdot, \cdot \}$, a vector field, $X_f$, is said
  to be its \textbf{Hamiltonian vector field} if 
$$
X_f(g)=\{f,g\}, \qquad \forall g\in C^\infty(M_C).
$$
In standard coordinates:
\begin{equation}
    X_f=\sum_{k=1}^n \frac{\partial f(R,P)}{\partial P_k}\frac{\partial}{\partial R^k}-
\frac{\partial f(R,P)}{\partial R^k}\frac{\partial}{\partial P_k}.
\end{equation}

We shall call a \textbf{Hamiltonian system} to a triple $(M_C,
\{\cdot, \cdot\}, H)$, where $\{ \cdot, \cdot\} $ is a Poisson bracket
on $M_C$, and $H\in C^\infty(M_C)$ is the Hamiltonian of the sytem.
The dynamics of a Hamiltonian system can be formulated in two
alternative manners:



\begin{itemize}
 \item The trajectories of the system are given by the integral curves of the Hamiltonian vector field
       $X_H$.
 \item If we consider instead the set of classical observables,
       the dynamics is written as the Poisson bracket of the
       Hamiltonian function $H$ with any other function, i.e.,
      \begin{equation}
        \label{eq:23}
        \frac{df}{dt}=\{ H, f\},\qquad \forall f\in C^\infty(M_C).
      \end{equation}
\end{itemize}
Both approaches are equivalent: the differential equations that determine
the integral curves of the Hamiltonian vector field are given by Eq.~(\ref{eq:23}), 
for the functions ``position'' and ``momenta'' of each particle (i.e., $\vec R$ and $\vec P$). These equations
are nothing else but Hamilton equations:
\begin{eqnarray}
  \label{eq:8}
\dot{R}^j& = & \displaystyle \frac{\partial H}{\partial P_j},\\
\dot{P}_j& = & \displaystyle -\frac{\partial H}{\partial R^j}.
\end{eqnarray}


\section{Summary of geometric quantum mechanics}
\label{sec:append-geom-quant}
The aim of this section is to provide a description of quantum
mechanical systems by using the same geometric tools which are used to
describe classical mechanical systems, outlined above. It is just a
very quick summary of the framework which has been developed in the
last 30 years and which can be found in
Refs.~\cite{kibble1979geometrization,heslot1985quantum,
  abbati1984pure,
  cirelli1991quantum,brody2001geometric,Ashtekar:1998p906,
  Carinena:2006p7565,Carinena:2007p813,clemente2008basics} and
references therein.
For the sake of
simplicity, we shall focus only on the finite dimensional case.
The Hilbert space $\mathcal{H}$ becomes then isomorphic to
$\mathbb{C}^n$ for $n$ a natural number, $\mathbb{C}^n\sim \mathcal{H}$:

\subsection{The states} 
Consider a basis $\{ |\psi_{k}\rangle\}$ for ${\cal H}$. 
Each state $|\psi\rangle\in {\cal H}$ can be written in that basis with complex components  
$\{ z_{k}\}$:
$$
\vert\psi\rangle=\sum_{k}z_{k}|\psi_{k}\rangle.
$$
We can just take the vector space inherent to the Hilbert
space, and turn it into a real vector space $M_Q$, by splitting each coordinate
in its real and imaginary part:
$$
z_{k}=q_{k}+ip_{k} \mapsto (q_{k}, p_{k})\in 
\mathbb{R}^{2n}\equiv M_{Q}.
$$
We will use real coordinates $(q_k, p_k)$, $k=1, \ldots ,n$ to
represent the points of $\mathcal{H}$ when thought as real manifold
elements. From this real point of view the similarities
between the quantum dynamics and the classical one described 
in the previous section will be more evident.
Sometimes it will be useful to maintain the complex notation $\psi$ 
or $z_k$ for the elements of the Hilbert space.    
Please notice that, despite the use of ``q'' and ``p'', these
coordinates have in principle no 
relation at all, with the position or the momentum of the quantum
system. They are 
simply the real and imaginary parts of the complex coordinates
used to represent the (finite-dimensional) Hilbert space vector in the
chosen basis.  Therefore, the resulting geometrical description has no
relation with the usual phase-space descriptions of Quantum Mechanics,
such as the Wigner representation, which is defined for
infinite-dimensional quantum systems and where the coordinates to
represent the point of the phase space do correspond to the spectra of
the position and momentum operators.  For the same reason, there is
no relation with any classical limit of the quantum system: the
description in terms of these coordinates $(q,p)$ is purely quantum.

There is no one-to-one correspondence   between physical states and
elements of the Hilbert space used to describe it; it is a well known
fact that physical states are independent of global phases.  This
ambiguity also translates to the formulation with real vector spaces.
Two equivalent states are related by a phase transformation; the best way
to characterize these transformations in $M_{Q}$ is by introducing
their infinitesimal generator, given by:
\begin{equation}
  \label{eq:phase}
  \Gamma := 
  \sum_{k}\left ( q_{k}\frac {\partial}{\partial p_{k}} -p_{k}
  \frac {\partial}{\partial q_{k}}\right ),
\end{equation}
This is a vector field whose meaning is clear if we realize that 
a phase change modifies the angle of the complex number representing the state,
when  considered in polar form 
(i.e., in polar coordinates $\{r_k,
\theta_k\}_{k=1, \cdots , n}$ with $z_k=r_k{\rm e}^{i\theta_k}$, Eq. 
(\ref{eq:phase}) becomes
$\Gamma=\sum_k\partial_{\theta_k}$). 
Then, from a geometrical point of view
we can use Eq. (\ref{eq:phase}) in two ways: 
\begin{itemize}
  \item Computing its integral curves, which are the different states
    which are obtained from an initial one by a global phase
    multiplication.
  \item Acting with the vector field on functions of $M_Q$. A function will be acceptable as an observable
    if it is invariant under phase transformations, i.e., $\Gamma f = 0$.
\end{itemize}

In a quantum Hilbert space, we also consider two states to be
equivalent if they merely differ by their norm. One may therefore only
consider the sphere of states with norm equal to one in
$\mathbb{C}^n$; this corresponds in the real-vector-space description
to the $(2N_Q-1)$--dimensional sphere:
\begin{equation}
  \label{eq:32}
  S_Q:=\left \{ (\vec q, \vec p)\in M_Q | \sum_k(q_k^2+p_k^2)=1
  \right \}. 
\end{equation}
It is immediate that the vector field $\Gamma$ is tangent to
$S_Q$, since the phase change preserves the norm of the state.



\subsection{The Poisson  bracket }


Taking as coordinates $z_k=q_k+ip_k$, the components of a vector in an
orthonormal basis we can define a Poisson bracket in $M_Q$, by:
\begin{equation}
\label{eq:poisson2}
\{f,g\}:=\sum_k \frac 12 \left(
\frac {\partial f}{\partial p_{k}}\frac {\partial g}{\partial q_{k}}
-
\frac {\partial f}{\partial q_{k}}\frac {\partial g}{\partial p_{k}}
\right),
\end{equation}
that corresponds to the standard Poisson bracket in classical mechanics.
It will also be useful to introduce a symmetric bracket by
\begin{equation}
\label{eq:symm}
 \{f,g\}_+:=
\sum_k \frac 12 
\left(\frac{\partial f}{\partial q_{k}}\frac {\partial g}{\partial q_{k}}
+
\frac {\partial f}{\partial p_{k}}\frac {\partial g}{\partial p_{k}}\right).
\end{equation}
We have made use of a specific basis; one may think that the
definition of the brackets could depend on the choice of the
basis. However, it is a matter of simple algebra to prove that this is
not the case.



An important property of the brackets defined above is that
they are preserved by the vector field $\Gamma$ in
Eq. (\ref{eq:phase}) in the sense that
$$
\Gamma\{f,g\}= \{\Gamma f,g\}+\{f,\Gamma g\}
$$
and
$$
\Gamma\{f,g\}_+= \{\Gamma f,g\}_++\{f,\Gamma g\}_+.
$$ A fact that can be proved with a simple computation. This property
is important for us because it implies that the symmetric or
antisymmetric bracket of two phase invariant functions will also be phase invariant:
$$
\Gamma f=\Gamma g=0\Rightarrow 
\Gamma\{f,g\}
=
\Gamma\{f,g\}_+=0.
$$

\subsection{The observables}
\label{sec:observ}

We now proceed to discuss how to represent the physical
observables on this new setting. Instead of considering the
observables as linear operators (plus the usual requirements,
self-adjointness, boundedness, etc.) on the Hilbert space
$\mathcal{H}$, we shall be representing them as functions defined on
the real manifold $M_Q$, as it is done in classical mechanics. But we
can not forget the linearity of the operators, and thus not any
function is acceptable. We will consider, for 
any operator $A\in \mathrm{Lin}(\mathcal{H})$, the quadratic function
  \begin{equation}
    \label{eq:2}
    f_A(\psi):=\langle\psi|A\psi\rangle. 
\end{equation}
We shall denote  the set of such functions as $\mathcal{F}_s(M_Q)$.


Notice that this definition of observable is different from
the analogous one in the classical case. In the classical framework a
state, represented by a point in $M_C$, provides a well defined result
for any observable $f:M_C\to \mathbb{R}$. In this  geometric quantum mechanics 
(GQM in the following) framework, on the
other hand, the value at a given state of  $f_A\in
\mathcal{F}_s(M_Q)$ provides just the average value of the corresponding
operator in that state. Besides, contrarily to the classical case,
not all the $C^\infty$ functions on $M_Q$ are regarded as observables, but
only those of the form of (\ref{eq:2}).

Once the definition has been stated, one must verify that it is a
consistent one, meaning that (i) they are phase invariant functions
($\Gamma f = 0$), and (ii) the set of observables is closed with
respect to both the symmetric and antisymmetric brackets (i.e. the
bracket of two elements of $\mathcal{F}_s(M_Q)$ is also an element of
$\mathcal{F}_s(M_Q)$). Both facts can be proven by direct computation.
The second proof, however, is interesting because it leads us to consider
the relevant algebraic operations within $\mathcal{F}_s(M_Q)$.
In the usual approach of quantum mechanics, there
are three algebraic structures on the set of operators, which turn
out to be meaningful and important for the physical description:
\begin{itemize}
\item The associative product of two operators: 
$$
A, B\in  \mathrm{Lin}(\mathcal{H}) \to A\cdot B \in \mathrm{Lin}(\mathcal{H}).
$$
It is important to notice, though, that this operation is not internal in
the set of Hermitian operators (i.e., those associated to physical magnitudes), since the 
product of two Hermitian operators is not Hermitian, in general. 
\item The anticommutator of two operators:
$$
A, B\in  \mathrm{Lin}(\mathcal{H}) \to [A,B]_+ \in \mathrm{Lin}(\mathcal{H}).
$$
\item The commutator of two operators:
$$
A, B\in  \mathrm{Lin}(\mathcal{H}) \to i[A,B] \in \mathrm{Lin}(\mathcal{H}).
$$
\end{itemize}
Notice that these last two operations are internal in the space of Hermitian operators.
How do we translate these operations into the GQM scheme? 
If we take  $f_A, f_B\in  \mathcal{F}_s(M_Q)$,
\begin{itemize}
\item the anticommutator of two operators, becomes the symmetric bracket
or Jordan product of the functions 
(see Ref. \cite{landsnab}):
$$
[A, B]_{+}\to f_{[A,B]_+}=\{f_A, f_B\}_+ \in \mathcal{F}_s(M_Q).
$$
\item The commutator of two operators, transkates into the Poisson bracket of
  the functions:
$$
i[A,B]\to f_{i[A,B]}=\{f_A, f_B\} \in \mathcal{F}_s(M_Q).
$$
\end{itemize}
We may conclude, therefore, that the operators and their algebraic
structures are encoded in the set of functions with the brackets
associated to the Hermitian product. Thus we see why it makes sense to
consider only that specific type of functions: it is a choice which
guarantees to maintain all the algebraic structures which are required
in the quantum description.

Another important property of the set of operators of quantum
mechanics is the corresponding spectral theory.  In any quantum
system, it is of the utmost importance to be able to find eigenvalues
and eigenvectors. In GQM, these objects are considered in the following way:
Let $f_{A}$ be the function associated to the observable $A$. Then, if we consider 
$\psi\in S_Q$,
\begin{itemize}
  \item the eigenvectors of the operator $A$ coincide with the critical points of the function $f_{A}$, i.e.,
$df_{A}(\psi)=0 \Rightarrow \psi$ is an eigenvector of $A$.

  \item The eigenvalue of $A$ at the eigenvector $\psi$ is the value that the function $f_{A}$ takes at the critical point $\psi$.
\end{itemize}

We finalize this discussion on the observables by noting that
the identity operator $\mathbb{I}$ corresponds with the function
\begin{equation}
f_\mathbb{I}(q,p)=\sum_{k}\left(q_{k}^{2}+p_{k}^{2}\right ),
\end{equation}
and the Hamiltonian vector field of this function is precisely 
the generator of phase transformations:
\begin{equation}
\{ f_\mathbb{I}, \cdot \}=\sum_k \left (p_{k}\frac{\partial} 
{\partial q_{k}}- q_{k}\frac{\partial}{\partial p_{k}}\right )=\Gamma.
\end{equation}


\subsection{The dynamics}

As in the classical case, the dynamics can be implemented in different forms, always in a way
which is compatible with the geometric structures introduced so far:
\begin{itemize}
\item  In the Schr{\"{o}}dinger picture, the dynamics is described as the integral curves
  of the vector field $X_{f_H}$, where $H$ is the Hamiltonian operator of the system:
  \begin{equation}
    \label{eq:4}
    X_{f_H}=\hbar^{-1}\{  f_H, \cdot \}.
  \end{equation}


\item In the Heisenberg picture, the dynamics is introduced by
  transferring the Heisenberg equation into the language of functions:
  \begin{equation}
    \label{eq:3}
    \dot f_A=\hbar^{-1}\{f_H, f_A \}.
  \end{equation}
  For example, the von Neumann equation is rewritten as:
\begin{equation}
  \label{eq:5}
   \dot f_\rho=\hbar^{-1}\{ f_H, f_\rho \},
\end{equation}
where $f_\rho$ is the function associated to the density matrix.

\end{itemize}
Therefore, it is possible to describe quantum dynamics as a
flow of a vector field on the manifold $M_Q$ or on the set of quadratic
functions of the manifold. Such a vector field is a Hamiltonian vector
field with respect to the Poisson bracket encoded in the
Hermitian structure of the system.

The dynamics thus defined preserves the norm of the state, which allows 
us to restrict the set of admissible states to those with norm equal to one.
This can be proven by analyzing the evolution of the function
$f_{\mathbb{I}}$, given by the Poisson bracket with the Hamiltonian: 
\begin{equation}
\dot f_{\mathbb{I}}=\hbar ^{-1}\{ f_{H}, f_{\mathbb{I}} \}.
\end{equation}
But because of the properties of the bracket, $\{f_H , f_{\mathbb{I}} \}=f_{[H,\mathbb{I}]}=0$,
and thus the flow is restricted to the sphere $S_Q$.

\begin{example}

\label{sec:dynamics}
  Let us consider a simple two-level quantum system, defined on
  $\mathbb{C}^2$. As a real manifold, $M_Q\sim \mathbb{R}^4$. Consider
  then a Hamiltonian $H:\mathbb{C}^2\to\mathbb{C}^2$:
\begin{equation}
H=\left[
\begin{array}{cc}
  H_{11} & H_{12} \\
H_{21} & H_{22}
\end{array}\right].
\end{equation}
If we consider it as a matrix on the real vector space $M_Q$, it
reads:
\begin{equation}
H_{\mathbb{R}}=\left[
\begin{array}{cccc}
  H_{q_1q_1} & H_{q_1p_1} &H_{q_1q_2} & H_{q_1p_2}\\
  H_{p_1q_1} & H_{p_1p_1} &H_{p_1q_2} & H_{p_1p_2}\\
  H_{q_2q_1} & H_{q_2p_1} &H_{q_2q_2} & H_{q_2p_2}\\
  H_{p_2q_1} & H_{p_2p_1} &H_{p_2q_2} & H_{p_2p_2}
\end{array}\right].
\end{equation}
It is easy to prove that this matrix is symmetric if the Hamiltonian is Hermitian ($H_{12} = H_{21}^*$).

The function $f_H$ in $\mathcal{F}_s(M_Q)$ becomes thus:
$$
f_H=\psi_{\mathbb{R}}^{t}H_{\mathbb{R}}\psi_{\mathbb{R}},
\quad{\mathit where}\quad \psi_{\mathbb{R}}=(q_{1},p_1,q_{2},p_2)^{t}
$$
and then, the Hamiltonian vector field turns out to be:
$$
  X_H=\hbar^{-1}\left ( \frac{\partial f_H}{\partial p_1}\frac{\partial}{\partial q_{1}}
- \frac{\partial f_H}{\partial q_{1}}\frac{\partial}{\partial p_1}+
\frac{\partial f_H}{\partial p_2}\frac{\partial}{\partial q_{2}}
-\frac{\partial f_H}{\partial q_{2}}\frac{\partial}{\partial p_2} \right )
$$
and its integral curves are precisely the expression of Schr\"odinger
equation when we write it back in complex terms:
\begin{eqnarray*}
    \dot q_{1}&=&\hbar^{-1}(H_{p_1q_1}q_{1}+H_{p_1q_2}q_{2}+H_{p_1p_1}p_1+H_{p_1p_2}p_2), 
    \\ 
    \dot p_1&=-&\hbar^{-1}(H_{q_1q_1}q_{1}+H_{q_1q_2}q_{2}+H_{q_1p_1}p_1+H_{q_1p_2}p_2),    
    \\
    \dot q_{2}&=&\hbar^{-1}(H_{p_2q_2}q_{2}+H_{p_2q_1}q_{1}+H_{p_2p_2}p_2+H_{p_2p_1}p_1),
    \\ 
    \dot p_2&=-&\hbar^{-1}(H_{q_2q_2}q_{2}+H_{q_2q_1}q_{1}+H_{q_2p_2}p_2+H_{q_2p_1}p_1).
\end{eqnarray*}


We can write these equations as:
\begin{equation}
  \label{eq:realsch}
\dot{\psi_{\mathbb{R}}}=-\hbar^{-1} \mathbf{J}
H_{\mathbb{R}}\psi_{\mathbb{R}},
\qquad
\mathit{with}
\qquad
\mathbf{J}= \left[
   \begin{array}{cccc}
     0 & -1 & 0 &  0 \\
     1 & 0 & 0 &0 \\
     0& 0 & 0 & -1 \\
     0 & 0 & 1 & 0
  \end{array}\right]
\end{equation}
or, equivalently, 
\begin{equation}
  \label{eq:7}
 |\dot\psi\rangle=-\hbar^{-1} i H|\psi\rangle,  
\qquad
\mathit{where}
\qquad
  |\psi\rangle=
\left[\begin{array}{c}
 q_{1} + i p_1 \\
 q_{2} + i p_2 
\end{array}\right]
\end{equation}
is the complex state vector in terms of the real coordinates. 
The operator $\mathbf{J}$, that satisfies $\mathbf{J}^2=-I$,
is called the complex structure. As it is apparent when comparing
(\ref{eq:realsch}) and (\ref{eq:7}), it
implements the multiplication by the imaginary unity $i$ 
in the real presentation of the Hilbert space. 
Observe that $\mathbf{J}$ and $H_{\mathbb{R}}$ commute, this is due to the
fact that the latter comes from an operator $H$ in the complex Hilbert space.
\end{example}

\section{Ehrenfest dynamics as a Hamiltonian system}
\label{sec:our-proposal}
In this section we show how to put together 
the dynamics of a quantum and a classical system 
following the presentation of the two previous sections.
We are describing thus a physical system characterized by the
following elements:

\subsection{The set of states of our system}
\begin{itemize}
 \item First, let $\mathcal{H}$ be  a Hilbert space  which
   describes the quantum degrees of freedom of our system.
   For example, it could describe the electronic subsystem; in this case, 
   it is the vector space corresponding to the completely
   antisymmetric representation of the permutation group $S_{N}$
   (i.e., a set of Slater determinants), where $N$ is the number of
   electrons of the system and each electron lives in a Hilbert space
   of dimension $M$. Thus, the dimension of ${\cal H}$ will be
   $N_{Q}=\frac{M!}{N!(M-N)!}$.

   We know that it is a complex vector space, but we prefer to consider it
   as a real vector space with the double of degrees of freedom and
   denote it as $M_Q$. Also, in correspondence with the Hilbert space
   vectors in the usual formalism of quantum mechanics, several states
   in $M_Q$ represent the same physical state. To consider true physical states one should 
   extract only those corresponding to the projective space, which can
   be identified with a submanifold of $M_Q$. A more general approach
   is to consider the sphere of states with norm equal to one, $S_Q$, and
   take into account  the phase transformations generated by Eq
   (\ref{eq:phase}) in a proper way. We will discuss this in the following sections.

 \item Second, let $M_C$ be  a differentiable manifold which
   contains the classical degrees of freedom. We shall assume it to
   be a phase space, and thus it will have an even number of degrees of
   freedom and it shall be endowed with a non degenerate Poisson bracket
   that in Darboux cordinates $(\vec R,\vec P)$ reads
   $$\{f,g\}_C=\sum_J \frac{\partial f}{\partial P_J}\frac{\partial g}{\partial R_J}-\frac{\partial f}{\partial R_J}\frac{\partial g}{\partial P_J}.$$

 \item  Third, we let our state space $\mathcal{S}$ be the Cartesian
   product of both manifolds, 
   $$
   \mathcal{S}:=M_C\times M_{Q}.
   $$
   Such a description has important implications: it is possible to consider each subsystem
   separately in a proper way but it is not possible to entangle the
   subsystems with one another. As long as Ehrenfest dynamics
   disregards this possibility, the choice of the Cartesian product is
   the most natural one.
\end{itemize}


The coordinates that describe the state of the system are:
\begin{itemize}
\item The positions and momenta of the nuclei: 
  \begin{equation}
    \label{eq:11}
    (\vec R, \vec P)\in M_C.
  \end{equation}
We will have $3N_C+3N_C$ of these, for $N_C$ the number of nuclei of the
system.  
\item The real and imaginary parts of the coordinates of the Hilbert
  space vectors with respect to some basis:
  \begin{equation}
    \label{eq:12}
    (\vec q, \vec p)\in M_Q.
  \end{equation}
We will have $N_Q+N_Q$ of these, for $N_Q$ the complex dimension of the
Hilbert space ${\cal H}$.
\end{itemize}

\subsection{The observables}

Our observables are functions defined on the state space
$S=M_C\times M_Q$. 
We know from our discussion in the case of a purely quantum system
that any function of the  form  (\ref{eq:2}) produces an evolution,
via the Poisson bracket, which preserves the norm. In the MQCD case, we
can easily write the analogue of the vector field (\ref{eq:phase}) by
writting: 
\begin{equation}
  \label{eq:17}
  \Gamma_{Q}=\mathbb{I}\otimes \Gamma.
\end{equation}

This is again the infinitesimal generator of phase transformations for
the quantum subsystem, but written at the level of the global state
space $M_{C}\times M_{Q}$.  A reasonable property to be asked to the
functions chosen to represent our observables is to be
constant under this transformation. From a mathematical point of view,
we shall define the set of possible observables, $\mathcal{O}$, as the set of all 
$C^\infty$--functions on the set $M_C\times M_Q$ which are constant
under phase changes on the quantum degrees of freedom, i.e., 
\begin{equation}
  \label{eq:10}
  \mathcal{O}:=\{ f\in C^\infty (M_C\times M_Q) | \, \, \Gamma_{Q}f=0 \}.
\end{equation}
As we shall see later, this choice reflects the fact that, when
considered coupled together, the nonlinearity of classical mechanics
expands also to MQCD.  There are some important subsets that should be
considered:
\begin{itemize}
\item The set of classical functions: these are functions which depend
  only on the classical degrees of freedom. Mathematically, they can be
  written as those functions $f\in \mathcal{O}$ such that there exists
  a function $f_C\in C^\infty(M_C)$ such that
$$
f(\vec R, \vec P, \vec q, \vec
p)=f_C(\vec R, \vec P)\,.
$$
We  denote
this subset as $\mathcal{O}_C$. An example of a function belonging
to this set is the linear momentum of the nuclei.

\item The set of \textbf{generalized quantum functions}: functions which depend
  only on the quantum degrees of freedom and which are constant under
  changes in the global phase.  Mathematically, they can be 
  written as those functions $f\in \mathcal{O}$ such that there exists
  a function $f_Q\in C^\infty(M_Q)$ such that
  \begin{equation}
    \label{eq:15a}
    f(\vec R, \vec P, \vec q, \vec
p)=f_Q(\vec q, \vec p); \quad \Gamma(f_{Q})=0\,.
  \end{equation}
We  denote these functions as $\mathcal{O}_Q$.  We have added
the adjective ``generalized'' because this set is too large to
represent only the set of pure quantum observables. These later functions, should
be considered, when necessary, as a smaller subset, which corresponds to
the set of functions defined in Eq. (\ref{eq:2}). We
denote this smaller subset as $\mathcal{O}_{Q}^s$. An example of
a function belonging to $\mathcal{O}_{Q}^s$ is the linear momentum of the electrons. 
\item A third interesting subset is the set of arbitrary linear
  combinations of the subsets above, i.e., those functions which are
  written as  the sum of a purely classical function and a purely
  quantum one:
\begin{eqnarray}
  \label{eq:15b}
  f(\vec R, \vec P, \vec q, \vec
  p)=&f_C(\vec R, \vec P)+f_Q(\vec q, \vec p).
\end{eqnarray}
We will denote this set as $\mathcal{O}_{C+Q}$. An element of this set
of functions is the total linear momentum of the composed system.
\end{itemize}

We would like to make a final but very important remark. We have not
chosen the set of observables as
\begin{equation}{\cal P}:=
\label{eq:42}
 \left \{ f_{A}\in C^\infty(M_C\times M_Q) |
    f_{A}=\langle\psi|
    A(\vec R, \vec P) \psi\rangle \right \}, 
\end{equation}
for $A(\vec R, \vec P)$ a linear operator on the Hilbert space
$\mathcal{H}$ depending on the classical degrees of freedom because of
two reasons:
\begin{itemize}
\item It is evident that the set above is a subset of (\ref{eq:10})
  and thus we are not loosing any of these observables. 
But it is a well known property that Ehrenfest dynamics is
  not linear it does not preserve $\mathcal{P}$. We must
  thus enlarge the set (\ref{eq:42}). 
\item  We are going to introduce in
  the next section a Poisson bracket on the space of observables. For
  that bracket to close a Poisson algebra, we need to consider the
  whole set (\ref{eq:10}). 
\end{itemize}

It is important to notice that in the set (\ref{eq:10}) there are
operators which are not representing linear operators for the quantum
part of the system and hence the set of properties listed in Section
\ref{sec:append-geom-quant} for the pure quantum case are meaningless
for them. But this is a natural feature of the dynamics we are
considering, because of its nonlinear nature.

\subsection{Geometry and the Poisson bracket on the classical-quantum world} 
\label{sec:dynam-class-quant}

As we assume that both the classical and the quantum subsystems are
endowed with Poisson brackets, we face the same problem we have when
combining, from a classical mechanics perspective, two classical
systems.  Therefore it is immediate to conclude that the corresponding
Poisson structures can be combined as:
\begin{equation}
  \label{eq:20}
\{ \cdot, \cdot\}:= \{ \cdot, \cdot\}_C+\hbar^{-1}\{ \cdot, \cdot\}_Q,
\end{equation}
where the term $ \{ \cdot, \cdot\}_C$ acts on the degrees of freedom
of the first manifold and  $\{ \cdot, \cdot\}_Q$ acts on the degrees
of freedom of the second one. It is a known fact \cite{Abraham:1978p1131} that 
such a superposition of Poisson brackets 
always produces a well defined Poisson 
structure in the product space.

The set of pure classical functions $\mathcal{O}_C$ and the set of
quantum generalized functions $\mathcal{O}_Q$  are closed under the Poisson
bracket.  The same happens with the quantum functions
$\mathcal{O}_Q^s$ and the set of linear combinations
$\mathcal{O}_{C+Q}$. In mathematical terms, what we have is a family of
Poisson subalgebras. This property ensures that the description of
purely classical or purely quantum systems, or even both systems at
once but uncoupled to each other, can be done within the formalism. 

Once the Poisson bracket on $M_{C}\times M_{Q}$ has been introduced,
one can prove, in analogy to the full quantum case, that the infinitesimal
generator of the phase transformations is the Hamiltonian vector field
of the identity operator, $\Gamma_Q = \{ f_{\mathbb{I}},\cdot\}$. Therefore,
a function is a legitimate observable if it commutes with $f_{\mathbb{I}}$:
$\{ f_{\mathbb{I}}, f\}=0$.


\subsection{The definition of the dynamics}

Analogously to the description given in Sections
\ref{sec:append-geom-dynam-PB} and \ref{sec:append-geom-quant}, the
dynamics of the mixed quantum classical systems can be implemented on
\begin{itemize}
\item The manifold which represents the set of states by defining a
  vector field whose integral curves represent the solutions of the
  dynamics (Schr\"odinger picture).
\item The set of functions (please note the differences between the
  classical and the quantum cases) defined on the set of states which
  represent the set of observables of the system. In this case the
  Poisson bracket of the functions with the Hamiltonian of the system
  defines the corresponding evolution (Heisenberg picture).
\end{itemize}

Both approaches are not disconnected, since they can be easily
related:
\begin{equation}
  \label{eq:18}
  X_H=\{ f_H,\cdot \}, 
\end{equation}
where we denote by $X_H$ the vector field which represents the
dynamics on the phase space and by $f_H$ the function which
corresponds to the Hamiltonian of the complete system.

We can now proceed to our first goal: to provide a Hamiltonian
description of Ehrenfest dynamics in terms of a Poisson structure. 
We thus define the following Hamiltonian system: 
\begin{itemize}
  \item A state space corresponding to the Cartesian product
        $M_C\times M_Q$.

  \item A set of operators corresponding to the set of functions
        $\mathcal{O}$ defined in Eq. (\ref{eq:10}).
        
  \item The Poisson bracket  defined in Eq. (\ref{eq:20}).
        
  \item And finally, the dynamics introduced by the following
        Hamiltonian function: 
        \begin{equation}
         \label{eq:9}
         f_H(\vec R, \vec P,\vec q, \vec p):=
         \sum_{J}\frac{\vec P_{J}^2}{2M_{J}}+\langle
         \psi(\vec q, \vec p)| H_e( \vec R) \psi(\vec q, \vec p)\rangle,
        \end{equation}
        where $H_e$ is the expression of the electronic Hamiltonian,  $M_J$
        are the masses of the classical subsystem of the nuclei and $\psi( \vec
        q, \vec p)$ is the real-space representation of the state $\psi$
        analogous to $|\psi\rangle$ in Eq. (\ref{eq:7}).        
\end{itemize}

 As a result, the dynamics of both subsystems are
obtained easily. In the
Schr\"odinger picture we obtain: 
\begin{eqnarray}
  \label{eq:6}
\dot{\vec R}&= \frac{\partial f_H}{\partial \vec P}=M^{-1}\vec P, 
\\
  \label{eq:6-2}
\dot{\vec P}&= -\frac{\partial f_H}{\partial \vec R}=-
\mathrm{grad}(\langle \psi(\vec q,\vec p)| H_e(\vec R) \psi(\vec q,
\vec p)\rangle), 
\\ 
  \label{eq:6-3}
\dot q_k&= \hbar^{-1}\frac{\partial f_H}{\partial p_k}, \quad
  k=1,\ldots,N_Q, 
\\
  \label{eq:6-4}
\dot p_k&= - \hbar^{-1}\frac{\partial f_H}{\partial q_k}, \quad
  k=1,\ldots,N_Q.
\end{eqnarray}
This set of equations corresponds exactly with Ehrenfest dynamics.

Finally, it is important to verify that 
the dynamics preserves the set of observables $\mathcal{O}$,
i.e., for any observable $f$, $\{f_H,f\} \in \mathcal{O}$.
This can be easily proven since we have established that
an observable belongs to $\mathcal{O}$ if it Poisson-commutes with 
  $f_{\mathbb{I}}$. Thus, as $f_H\in \mathcal{O}$, if we consider an
  observable $f\in \mathcal{O}$,  by the Jacobi identity:
  \begin{equation}
\{ f_{\mathbb{I}},\{f_H, f\}\} =-\{f,\{f_{\mathbb{I}},
    f_H\}\}-\{f_H,\{f, f_{\mathbb{I}}\}\}=0.
\end{equation}


\section{A phase space description of the statistics of the 
Ehrenfest dynamics}
\label{sec:defin-stat-syst}

The next step in our work is  the definition of a statistical system
associated to the dynamics we introduced above. The first ingredient
for that is the definition of the distributions we shall be describing
the system with.

The main conclusion from the previous section is that Ehrenfest dynamics
can be described as a Hamiltonian system on a Poisson
manifold. Therefore, we are in a situation similar to a standard
classical system.  We have seen that the dynamics preserves the
submanifold $M_C\times S_Q$ ($S_{Q}$ being the sphere  in Eq. (\ref{eq:32})),
and thus it is natural to consider such a manifold as the space of our
statistical system. 

We have thus to construct now a statistical system composed of two
subsystems. We may think in introducing a total distribution 
factorizing as the product of a classical and a quantum distribution. But as the
events of the classical and the quantum regimes are not independent,
the two probabilities must be combined and not defined through 
two factorizing functions.

Hence we shall consider a density $F_{QC}\in
\mathcal{O}$ and the canonical volume element $d\mu_{QC}$ in the
quantum-classical phase space, restricted to $M_C\times S_Q$:
\begin{equation}
\label{eq:volumeform}
d\mu_{QC}:=d\mu_Cd\Omega_Q\,,
\end{equation}
which allows to define the macrosopic average for any observable $M\in\mathcal{O}$:
  \begin{equation}
    \label{eq:15}
    \langle M\rangle:=\int_{M_C\times S_Q}\hspace{-0.2cm}M(\vec R, \vec P, \vec q, \vec
    p) F_{QC}(\vec R, \vec P, \vec q, \vec p)d\mu_{QC}.
  \end{equation}
In the rest of the section we shall discuss in more detail the
ingredients of this definition.

We start by the volume form. Given the classical
phase space $M_C$ with Darboux coordinates $(\vec R,\vec P)$
(they always exist locally \cite{Abraham:1978p1131}) we define the volume 
element in the classical phase space by
\begin{equation}
\label{eq:volumeclassical}
d \mu_C:=\prod_JdP_JdR_J,
\end{equation}
which, as it is well known, is invariant under any purely classical
Hamiltonian evolution, and does not depend on the choice of coordinates.

We might proceed in the same way for the quantum part of our system.
Indeed, as we have written the quantum dynamics like a classical Hamiltonian one
we could use the canonical invariant
volume element describe in the previous paragraph $d\mu_Q$ on $M_Q$.
However, new complications appear in this case as in the quantum part
of the phase space we restrict the integration to $S_Q$ and then we look 
for an invariant volume form in the unit sphere, rather than in the full 
Hilbert space. 

The new ingredient that makes the restriction possible is the fact
that all our observables and of course, also the Hamiltonian,
are killed by $\Gamma$. This implies, 
as discussed before, that $f_{\mathbb I}$ is
a constant of motion.
Using this we decompose

\begin{equation}
  \label{eq:24}
d\mu_Q= df_{\mathbb I}d\widetilde\Omega_Q,  
\end{equation}
which is possible on ${\cal H}\setminus\{0\}$. 
The decomposition is not unique, but the restriction (pullback) 
of  $d\widetilde\Omega_Q$ to the unit sphere gives a uniquely defined, 
invariant volume form on $S_Q$ that we will denote by $d\Omega_Q$.
The decomposition is nothing but the factorization into the radial 
part and the solid angle volume element.

\begin{example}
  For the simple case we studied above, where
$\mathcal{H}=\mathbb{C}^2$ and hence $M_Q=\mathbb{R}^4$ and $S_Q$ is a
three dimensional sphere, the volume element above would be:
  \begin{equation}
    \label{eq:28b}
    d\Omega_{Q}=q_1dq_2dp_1dp_2-q_2dq_1dp_1dp_2 
               +p_1dq_1dq_2dp_2-p_2dq_1dq_2dp_1.
  \end{equation}
\end{example}

Now, we finally put together the two ingredients to obtain
the invariant volume element $d\mu_{QC}$ in eq. (\ref{eq:volumeform}).

We discuss now the properties we must require from $F_{QC}$ in order
for  Eq. (\ref{eq:15}) to correcly define the statistical mechanics for the
Ehrenfest dynamics.

\begin{itemize}
  \item The expected value of the constant observable should be that constant, 
        which implies that the integral on the whole set of states is equal to one:
        \begin{equation}
         \label{eq:16}
         \int_{M_C\times S_Q}F_{QC}d\mu_{QC}=1.
        \end{equation}
  \item The average, for any observable $f_A$ of the form (\ref{eq:42}) 
        associated to a positive definite hermitian operator $A$ should 
        be positive. This implies the usual requirements of positive 
        probability density in purely classical statistical mechanics.
\end{itemize}

It is simple to go from this phase space description  to the formulation of a
 density operator
$\rho$ for the quantum system and make the construction closer to the
usual description of Quantum Statistical Mechanics. The density
operator is defined as follows:
\begin{equation}
  \label{eq:13}
  \rho(\vec R, \vec P):=\int_{S_Q}d\mu_Q(\vec q, \vec p)F_{QC}(\vec R, \vec P, \vec q, \vec
  p) |\psi(\vec q, \vec p)\rangle\langle\psi(\vec q, \vec p)|,
\end{equation}
where $|\psi(\vec q, \vec p)\rangle\langle\psi(\vec q, \vec p)|$ is
the projector on the quantum state parametrized by the pair $(\vec q,
\vec p)$. The average value of a pure quantum magnitude $A$, represented by the 
quadratic funtion $f_{A}$ can be
computed as:
\begin{equation}
  \label{eq:21}
  \langle A\rangle (\vec R, \vec P):= \mathrm{Tr} (\rho(\vec R, \vec
  P) A)=
\int_{S_Q}d\mu_Q(\vec q, \vec p)F_{QC}(\vec R, \vec P, \vec q, \vec
p)f_A(\vec q, \vec p).
\end{equation}

To obtain the corresponding total average, it is required to integrate
the resulting object on the classical phase space:
\begin{equation}
  \label{eq:22}
  \langle A\rangle:=\int_{M_C}d\mu_C(\vec R, \vec P)\langle A\rangle (\vec R, \vec P).
\end{equation}

So far we have discussed general considerations about the statistical 
mechanics of our system. Now we should include the dynamics in the game.
Given that we have formulated the Ehrenfest dynamics as a Hamiltonian
system, it is well known \cite{balescu1997statistical} which is the evolution 
of the density function $F_{QC}$. It is given by the {\bf Liouville equation}
\begin{equation}
  \label{eq:19}
  \frac{dF_{QC}}{dt}+\{f_H, F_{QC}\}=0,
\end{equation}
where, in the derivation of the equation, it is an essential requirement
the invariance of the volume form under the evolution of 
the system.

The equilibrium statistical mechanics is obtained by requiring $\dot F_{QC}=0$, 
which making use of (\ref{eq:19}) is equivalent to
$$\{f_H, F_{QC}\}=0,$$ or, in other words, $F_{QC}$ should be a
constant of motion.  An obvious non trivial constant of motion is
given by any function of $f_H$. The question is if there are more. To
answer this we should examine closely our particular dynamics. In
several occasions we have mentioned the non linear character of a
generic Ehrenfest system. This is an essential ingredient, because due
to this fact we do not expect any other constant of motion that a
function of the Hamiltonian itself.  This is in contrast with the case
of linear equations of motion, where the system is necessarily
integrable.

The question of the existence of more constants of motion and,
therefore, more candidates for the equilibrium density $F_{QC}$ can be
elucidated with a simple example that we discuss below.  In it we show
the presence of ergodic regions (open regions in the leave of constant
energy densely covered by a single orbit).  This rules out the
existence of constants of motion different from the constant function
in the region (i.e., functions of $f_H$).

\begin{example}
In the following example we will study a simple toy model
in which the coupling between classical and quantum degrees of freedom
gives rise to chaotic behavior and the appearence of ergodic regions.

The system consists of a complex two dimensional Hilbert space
$M_Q=\mathbb{C}^2$ and a classical 2-D phase space where we define a 1-D harmonic 
oscillator. Using coordinates $(I_{\theta},\theta)$ for the classical variables (action-angle 
coordinates for the oscillator)
and $\Psi\in\mathbb{C}^2$ we define the following Hamiltonian
$$f_H=I_{\theta} + {\frac 1 2}\langle
\Psi\vert\sigma_z
+\epsilon \cos(\theta)\sigma_x\vert\Psi\rangle,$$
with $\sigma_x, \sigma_z$ the Pauli sigma matrices. 

We parametrize the normalized quantum state by
\begin{equation}
\vert\Psi\rangle=e^{i\alpha}
\left[\begin{array}{c}
I_{\phi} \\ 
e^{i\phi}\sqrt{1-I_{\phi}^2}
\end{array}\right]\,.
\end{equation}
One can now easily check that $I_{\phi}$ and $\phi$ are canonical
conjugate variables. In these variables the Hamiltonian reads
\begin{equation}
  \label{eq:44}
  f_H= I_{\theta}+ I_{\phi}^2+{\epsilon}I_{\phi}\sqrt{1-I_{\phi}^2}
\cos(\theta)\cos(\phi),
\end{equation}
where $\epsilon$ measures the coupling of the classical and
quantum systems.

In the limit of vanishing $\epsilon$ the system
is integrable and actually linear in these coordinates.
However for non vanishing $\epsilon$ the model becomes non linear
and, as we will show below, regions of chaotic motion emerge.

In order to understand the behaviour of our system it is useful
to study its Poincar\'e map (see 
Ref. \cite{Abraham:1978p1131} for the definition). To this end we take the 
transversal (or Poincar\'e)  section at 
$\theta=0$ and, taking into account the conservation of energy,
only two coordinates are needed to describe the map, we have chosen
the quantum variables $I_{\phi}$ and $\phi$.\footnote{ In
our construction of the Poincar\'e map we represent the value of 
$\phi$ and $I_\phi$ whenever the periodic coordinate $\theta$
takes the value $\theta=0$. The successive values of
$\phi$ and $I_\phi$, for some initial condition,
form our orbit. 
As stressed in \cite{Abraham:1978p1131} the Poincar\'e map is a very 
useful tool in the context of non linear dynamics as many properties 
of the full system like ergodicity, chaos or regular behaviour, may be
inferred from the same property of the Poincar\'e map.}

\begin{figure}[t!]
  \centering
\includegraphics[width=7cm]{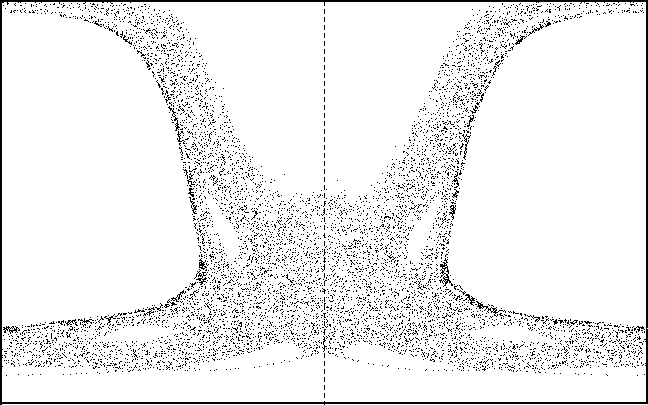}

  \caption{The plot shows a single orbit of the Poincar\'e map   of
Eq. (\ref{eq:44}) at $\theta=0$.  The 
angle $\phi\in[-\pi,\pi]$ is plotted in the horizontal axis and 
$I_{\phi}\in[0,1]$ in the vertical one.
We have taken $\epsilon=0.8$. The energy determines $I_{\theta}$  and its actual
value does not affect the dynamics of $\phi$ and $I_{\phi}$.}
\label{fig:1}
\end{figure}

By examining the plot of Fig. \ref{fig:1}, one sees that the orbit
densely fills a large region of the 
Poincar\'e section and therefore any constant of motion should be the
constant function in that region.
In the presence of such an ergodic evolution the only constants of motion
are functions of the Hamiltonian.
\end{example}

As a consequence we can claim that, generically within the Ehrenfest
dynamics, the only functions which commute with the Hamiltonian in
(\ref{eq:19}) are those which are function of the Hamiltonian $f_H$
itself. As the Poisson bracket is skew-symmetric, the property follows
trivially. Ergodicity helps us to assert, roughly speaking, that in
general no other function in $\mathcal{O}$ will be a constant of the
motion.

Thus, at this stage, we can finish our construction with the
equilibrium distribution associated to Ehrenfest dynamics. Taking into
account the ergodicity of the dynamics, we can impose the
equal-probabilities condition to the configurations of an isolated Ehrenfest system, which 
leads to the microcanonical ensemble, (see Ref. \cite{Oliveira2007ergodic}). 

In the case in which this system is weakly coupled to a bath, if we assume that the dynamics of the system plus bath is ergodic as well, such that system plus bath are microcanonically distributed, then the statistics of the Ehrenfest system are given by the well known canonical probability density  (see, for example, Refs. \cite{balescu1997statistical,balescu1975equilibrium,
  Schwabl2002statistical}):
\begin{equation}
  \label{eq:31}
  F_{QC}^{c}(\vec R, \vec P, \vec q, \vec p) = Z^{-1}e^{-\beta f_H},
  \qquad
  Z = \int_{M_C\times S_Q}e^{-\beta f_H}d\mu_{QC}\,,
\end{equation}
where $\beta^{-1}$, proportional to the temperature of the subsystem,
is the parameter governing the equilibrium between the system and the bath.
An information-theoretic approach to the
equiprobability in the microcanonical ensemble and to
Eq. (\ref{eq:31}) may be seen in Ref. \cite{Jaynes1957maxent1} or in
Refs. \cite{Ellis1985largedev,Touchette2009largedev} for a much more
modern and mathematically sound exposition.  We also strongly
recommend the recent work of P. Reimann
\cite{Reinmann2008foundations, Reinmann2010njp} for a foundation of
Statistical Mechanics under experimentally realistic conditions.

\section{Application: Nos\' e formalism for Ehrenfest systems}

\label{sec:application:-nos-e}

We now procced to extend the non-stochastic method proposed by
Nos\'e \cite{nose:1984,nose:1991} to our Hamiltonian
(\ref{eq:9}). Consider a system defined on $M_C\times M_Q$ as before and
an extension of that system including two new classical degrees of
freedom $(s, p_s)$, what produces a total state space
$$
M_C\times M_Q\times \mathbb{R}^2.
$$

In this case, in order to include the quantum degrees of freedom, we
must take into account the problem of the deformation of the domain
(see the change of coordinates of Eq.~(\ref{eq:21n})) of integration we
dealt with in the previous section. Thus we must rewrite the
Hamiltonian given in Eq.~(\ref{eq:9}) in order to extend the domain of
the quantum part from the sphere to the whole $M_Q$. In order to do
this, we consider:
 \begin{eqnarray}
    H(\vec R, \vec P, \vec q, \vec p) & := &\sum_{k}\frac{\vec
      P_k^2}{2M}+\frac{\langle\psi (\vec q, \vec p)|H^e(\vec R) \psi(\vec
    q, \vec p)\rangle}{\langle\psi(\vec q, \vec p)|\psi(\vec q, \vec
    p)\rangle} \nonumber
    \\    \label{eq:23n}
    & =: & H^N(\vec P)+e_{H^{e}}(\vec q, \vec p, \vec R),
  \end{eqnarray}
Now we transform this Hamiltonian into the ``extended'' one, by
introducing a dependence with respect to the new degrees of
freedom:
\begin{equation}
\label{eq:14b}
  H(\vec R, \vec P, \vec q, \vec p, s, p_s) := H^N(\vec P/s)+e_{H^{e}}(\vec q, \vec p/s, \vec R) 
  + \frac{p_s^2}{2Q}+gkT\log (s) \ .
\end{equation}

This is very similar to the method proposed by Nos\'e 
\cite{nose:1984,nose:1991} for thermostating classical systems. The Nos\'e
technique is basically an algorithm to produce the correct equilibrium
distribution and, therefore, the interpretation of the coordinate
$s$ as describing the behaviour of some `thermostat' is only metaphorical.
In the same sense, the particular form of the last two terms in the
above Hamiltonian is just the one needed for the theorem that we mention
in the following lines to work.

The extended equations of motion read now:
\begin{eqnarray}
\label{eq:1}
    \dot{\vec R}=\vec{P}/(Ms^2),
    \\
    \dot{\vec P}=-\mathrm{grad}_{\vec R} \left (e_{H^{e}}(\vec q, \vec p/s, \vec R) \right ), 
    \\
   \dot q_{k}=\frac{\hbar^{-1}s^{-1}}{\langle\psi |\psi\rangle}(H_{p_kq_1}q_{1}+\dots +H_{p_kp_{N_Q}}p_{N_Q}/s -e_{H^{e}}(\vec q, \vec p/s, \vec R)p_k), 
    \\ 
    \dot p_k=-\frac{\hbar^{-1}}{\langle\psi|\psi\rangle}(H_{q_kq_1}q_{1}+\dots +H_{q_kp_{N_Q}}p_{N_Q}/s-e_{H^{e}}(\vec q, \vec p/s, \vec R)q_k), 
    \\
  \dot s=\frac{p_s}Q, \\
\label{eq:14}
\dot p_s=s^{-1}\sum_k \dot q^k+\sum_j \dot R^j-s^{-1}gKT,
\end{eqnarray}
where $k=1, \dots, N_{Q}$.

Now we generalize the standard result of Nos\'e \cite{nose:1984,
  nose:1991}, by proving the equivalence of the microcanonical
distribution of the extended system and the canonical distribution of
the original one. The equations (\ref{eq:1}-\ref{eq:14}) are the necessary ones to
implement the microcanonical scheme once we assume the ergodic
hypothesis for the extended system. To prove the equivalence with the
canonical distribution of the original system, we take the Hamiltonian
above, and we consider the corresponding microcanonical distribution:
\begin{equation}
  \label{eq:19n}
  Z=\int_{M_C\times M_Q\times \mathbb{R}^2}d \mu_{C}d\mu_{Q}dp_sds\delta( H(\vec
  R, \vec P, \vec q,\vec p,s, p_{s})-E),
\end{equation}
where $d\mu_{C}d\mu_{Q}$ is the volume element on $M_C\times M_Q$.

If we consider the following noncanonical transformation  on
$M_C\times M_Q\times \mathbb{R}^2$ :
\begin{equation}
  \label{eq:21n}
R'^j=R^j, \quad
P'_j=  \frac{P_j}{s}, \quad
q'^k=q^k, \quad
 p'_k=\frac{p_k}{s}, \quad
 s'=s, \quad
 p_s'=p_s,
\end{equation}
the partition function reads (we use the same symbol again for the
transformed variables but include the jacobian which is equal to $s^D$):
\begin{eqnarray}
Z&=&\int_{M_C\times M_Q\times \mathbb{R}^2}d \mu_{C}d\mu_Qdp_Sds\,\,{s}^{D}
\nonumber \\
&&
\delta \left ( H^N(\vec P) + e_{H^e}({\vec q}, \vec p, \vec R )+
\frac{{p}_s^2}{2Q}+gkT\log (s)-E\right ),  
\end{eqnarray}
where $D$ is the number of degrees of freedom of the system (the sum of classical and 
quantum degrees of freedom).

As the original integral domain includes the complete $M_Q$, and
therefore a rescaling of momenta does not modify it, we can use the
equivalence
$$
\delta(f(s))=\delta(s-s_0)/|f'(s_0)|,
$$
where 
$$
s_0=\exp \left ( -\frac{H^N(\vec P)+e_{H^e}(\vec q, \vec p, \vec R)+ \frac{{p}_s^2}{2Q}-E}{gkT}\right ),
$$
and write:
\begin{eqnarray}
\nonumber
Z & = & \int_{M_C\times M_Q\times \mathbb{R}^2}d\mu_{C}d\mu_{Q}dp_Sds\,\,{s}^{D}
\\
& &  \delta \left [s -\exp \left ( -\frac{H^N(\vec P)+e_{H^e}(\vec q, \vec p, \vec R) + \frac{{p}_s^2}{2Q}-E}{gkT}\right )\right ].
\end{eqnarray}
Integrating in the variable $s$, we get:
\begin{eqnarray}
\nonumber
Z & = & \frac 1{gkT}\int dp_s \exp \left [ \frac{D+1}{gkT}\left
    (E-\frac{{p}_s^2}{2Q} \right )\right ]
\\
& & \int_{M_C\times M_Q}d \mu_{C}d\mu_{Q} \exp \left [-\frac{D+1}{gkT} (H^N(\vec
  P)+e_{H^e}(\vec q, \vec p, \vec R)\right ].
\end{eqnarray}

Then if we choose $g=D+1$ the exponential takes the desired
form. Besides, we can notice that when calculating the average
of  any function  $M\in \mathcal{O}$ (where $\mathcal{O}$ corresponds to
Eq. (\ref{eq:10}) and therefore $M$ does not depend on the extended
variables), the quantity:  
$$
\frac 1{(D+1)kT}\int dp_s \exp \left [ \frac{1}{kT}\left
    (E-\frac{{p}_s^2}{2Q} \right )\right ]
$$
factorizes.  If we also assume that the function $M$  does not depend either on the norm
of the quantum state (as it is the case for the quantum observables) and  we
decompose the volume element $d\mu_{Q}$ as Eq. (\ref{eq:24}),
the radial part of the quantum integral  also factorizes and cancels the analogous term 
arising from the partition function. 

Thus we can write
{\small 
\begin{eqnarray*}
  \label{eq:25}
  \langle M \rangle&=&  
1/Z\int_{M_C\times M_Q\times \mathbb{R}^2}\hspace{-0.4cm}d\mu_{C}df_{\mathbb{I}}
d\tilde \Omega_{Q}dp_sds M(\vec   R, \vec P, \vec q,\vec p) 
 \delta( H(\vec   R, \vec P, \vec q,\vec p,s, p_{s})-E) \\
&=& 
1/Z \left ( \frac 1{(D+1)kT}\int dp_s \exp \left [ \frac{1}{kT}\left
    (E-\frac{{p}_s^2}{2Q} \right )\right ]\int_{\mathbb{R}}df_{\mathbb{I}} \times \right . \\
&& \left  . \int_{M_C\times S_Q}\hspace{-0.4cm}d\mu_{C}
d\Omega_{Q}M(\vec   R, \vec P, \vec q,\vec p)  \exp \left [-\frac{1}{kT} \left (H^N(\vec
  P)+e_{H^e}(\vec q, \vec p, \vec R)\right )\right ] \right )
\end{eqnarray*}}
where 
{\small
\begin{eqnarray*}
Z&=&\frac 1{(D+1)kT}\int dp_s \exp \left [ \frac{1}{kT}\left
    (E-\frac{{p}_s^2}{2Q} \right )\right ]  \int_{\mathbb{R}}df_{\mathbb{I}} \times \\
&&\int_{M_C\times S_Q}d \mu_{C}
d\Omega_{Q} \exp \left [-\frac{1}{kT} \left (H^N(\vec
  P)+e_{H^e}(\vec q, \vec p, \vec R)\right )\right ].
  \end{eqnarray*}
}
 Hence we have proved that the microcanonical distribution of the extended dynamical system 
created by Eq.(\ref{eq:14b}) is equivalent to the canonical distribution created by the 
Hamiltonian
$$
H(\vec R, \vec P, \vec q, \vec p)=H^N(\vec  P)+e_{H^e}(\vec q, \vec p, \vec R).
$$

 
\section{Conclusions and future work}
\label{sec:concl-future-work}

In this paper we have constructed a rigorous Hamiltonian description
of the Ehrenfest dynamics of an isolated system by combining the
Poisson brackets formulation of classical mechanics with the geometric
formulation of quantum mechanics. We have also constructed the
corresponding statistical description and obtained the associated
Liouville equation.  Finally, after verifying numerically that the
Ehrenfest dynamics is ergodic, we justify the equilibrium distribution
produced by it.

It is important to keep in mind the restriction to finite-dimensional quantum
systems that we fixed as starting point, since our first goal is to apply our
construction to Molecular Dynamics. From a theoretical point of view, though,
it is very interesting to perform the same type of analysis for infinite
dimensional quantum systems (note that the Ehrenfest dynamics as formulated in
\cite{bornemann1996quantum,bornemann1995pre} consists of classical nuclei and
an infinite-dimensional Hilbert space for the electronic subsystem). Formally,
the construction can be defined in the same way, once we adapt the language to
infinite dimensional manifolds. But the nonlinearity of the resulting dynamics
imposes very challenging problems at many levels, for instance at the very
existence of the solutions of the Hamiltonian dynamical flow. Another
interesting problem  would be to recover, within the geometrical
formalism,  the classical-quantum system as  a suitable classical 
limit for the nuclear degrees of freedom of an original completely
quantum system (similarly to what is done in
\cite{bornemann1996quantum,bornemann1995pre} in the standard approach).  We hope to 
address these problems in the future.

The definition of the Hamiltonian in Eq.(\ref{eq:9}) and its associated
canonical equilibrium distribution Eq. (\ref{eq:31}) allows us now to use
(classical) Monte Carlo methods for computing canonical equilibrium averages
in our system, given by the expression Eq. (\ref{eq:15}). But this formalism also
allows to easily introduce a Nos{\'{e}} extended system scheme to compute
averages in the canonical ensemble by performing constant energy Molecular
Dynamics simulations. This is a very important feature of the final setting,
because the situations where the Ehrenfest approach is necessary (i.e. those
where the electronic energy gap is small and the Born-Oppenheimer
approximation is not sufficient) the effect of the temperature can be
relevant. A method for the simulation of the canonical ensemble is thus of the
greatest importance.

Alternatively, if we wish to perform MD simulations using stochastic
methods, we should first construct the Langevin dynamics associated
with our Hamiltonian Eq. (\ref{eq:9}) and equations
Eqs. (\ref{eq:6})-(\ref{eq:6-4}). To develop this program, we also need an
analogue of the Fokker-Planck equation associated with those Langevin
equations and then we have to check whether its solution at infinite
time approaches Eq.(\ref{eq:31}). Presently we are working on this point
\cite{futurepaper}, i.e., on the extension of our formalism to an
associated stochastic MD.

\section*{Acknowledgments}

We would like to thank Jos\'e F. Cari\~nena, Andr\'es Cruz and David
Zueco for many illuminating discussions. This work has been supported
by the research projects E24/1, E24/2 and E24/3 (DGA, Spain), FPA2009-09638
(CICYT, Spain), 
FIS2009-13364-C02-01 (MICINN, Spain), 200980I064 (CSIC, Spain)
and ARAID and Ibercaja grant for young researchers (Spain).
We also acknowledge the suggestions and insights of the referees.


\section*{References}


\begin{thebibliography}{10}

\bibitem{marxhutter:2009}
D.~Marx and J.~Hutter.
\newblock {\em Ab initio {M}olecular {D}ynamics: basic theory and advanced
  methods}.
\newblock Cambridge University Press, 2009.

\bibitem{gerber1982time}
R.~B. Gerber, V.~Buch, and M.~A. Ratner.
\newblock Time-dependent self-consistent field approximation for intramolecular
  energy transfer {I}. {F}ormulation and application to dissociation of van der
  {W}aals molecules.
\newblock {\em J. Chem. Phys.}, 77:3022-3030, 1982.

\bibitem{gerber1988self}
R.~B. Gerber and M.~A. Ratner.
\newblock  Self-Consistent-Field Methods for Vibrational Excitations in
  Polyatomic Systems, Advances in Chemical Physics 70, pages 97--132.



\bibitem{bornemann1996quantum}
F.~A. Bornemann, P.~Nettesheim, and C.~Sch{\"u}tte.
\newblock Quantum-classical molecular dynamics as an approximation to full
  quantum dynamics.
\newblock {\em J. Chem. Phys.}, 105(3):1074-1083, 1996.

\bibitem{bornemann1995pre}
F.A. Bornemann, P.~Nettesheim, and C.~Sch{\"u}tte.
\newblock Quantum-classical molecular dynamics as an approximation to full
  quantum dynamics.
\newblock Technical Report SC-95-26, Konrad-Zuse-Zentrum, 1995.

\bibitem{Zhu2005JCTC}
C.~Zhu, A.~W. Jasper, and D.~G. Truhlar.
\newblock Non-born-oppenheimer liouville-von neumann dynamics. evolution of a
  subsystem controlled by linear and population-driven decay of mixing with
  decoherent and coherent switching.
\newblock {\em J. Chem. Theor. Comp.}, 1:527-540, 2005.

\bibitem{heslot1985quantum}
A.~Heslot.
\newblock Quantum mechanics as a classical theory.
\newblock {\em Phys. Rev. D}, 31(6):1341--1348, 1985.

\bibitem{Par2006JCTC}
P.~V. Parandekar and J.~C. Tully.
\newblock Detailed balance in {Ehrenfest} mixed quantum-classical dynamics.
\newblock {\em J. Chem. Theor. Comp.}, 2:229-235, 2006.

\bibitem{Kab2006JPCA}
G.~K{\"{a}}b.
\newblock Fewest switches adiabatic surface hopping as applied to vibrational
  energy relaxation.
\newblock {\em J. Phys. Chem. A}, 110:3197-3215, 2006.

\bibitem{Bas2006CPL}
A.~Bastida, C.~Cruz, J.~Z{\'{u}}{\~{n}}iga, A.~Requena, and B.~Miguel.
\newblock A modified {Ehrenfest} method that achieves {Boltzmann} quantum state
  populations.
\newblock {\em Chem. Phys. Lett.}, 417:53-57, 2006.

\bibitem{Kab2002PRE}
G.~K{\"{a}}b.
\newblock Mean field {Ehrenfest} quantum/classical simulation of vibrational
  energy relaxation in a simple liquid.
\newblock {\em Phys. Rev. E}, 66:046117, 2002.

\bibitem{Tul1998Book}
J.~C. Tully.
\newblock {\em Nonadiabatic Dynamics}, pages 34--72.
\newblock World Scientific, Singapore, 1998.

\bibitem{Mul1997JCP}
U.~M{\"{u}}ller and G.~Stock.
\newblock Surface-hopping modeling of photoinduced relaxation dynamics on
  coupled potential-energy surfaces.
\newblock {\em J. Chem. Phys.}, 107:6230-6245, 1997.

\bibitem{Mau1993EPL}
F.~Mauri, R.~Car, and E.~Tosatti.
\newblock Canonical statistical averages of coupled quantum-classical systems.
\newblock {\em Europhys. Lett.}, 24:431-436, 1993.

\bibitem{Ter1991JCP}
T.~Terashima, M.~Shiga, and S.~Okazaki.
\newblock A mixed quantum-classical molecular dynamics study of vibrational
  relaxation of a molecule in solution.
\newblock {\em J. Chem. Phys.}, 114:5663-5673, 2001.


\bibitem{Alonso:2010p6480}
J.~L. Alonso, A.~Castro, P.~Echenique, V.~Polo, A.~Rubio, and D.~Zueco.
\newblock Ab initio molecular dynamics on the electronic {Boltzmann}
  equilibrium distribution.
\newblock {\em New J. Phys.}, 12:083064, 2010.


\bibitem{nose:1984}
S.~Nos{\'e}.
\newblock A unified formulation of constant temperature molecular dynamics
  methods.
\newblock J. Chem Phys. 81 (1) 511-519, 1984.

\bibitem{nose:1991}
S.~Nos{\'e}.
\newblock Constant temperature molecular dynamics methods.
\newblock {\em Prog. Theor. Phys. Suppl}, 103:1-46, 1991.

\bibitem{Kisil:fk}
V.~V. Kisil.
\newblock No more than mechanics. {I}.
arXiv:funct-an/9405002v3 and
\newblock {\em J. Natur. Geom.}, 9(1):1--14, 1996.

\bibitem{prezhdo1997mixing}
O.~V. Prezhdo and V.~V. Kisil.
\newblock Mixing quantum and classical mechanics.
\newblock {\em Phys. Rev. A}, 56(1):162--175, 1997.

\bibitem{kapral1999mixed}
R.~Kapral and G.~Ciccotti.
\newblock Mixed quantum-classical dynamics.
\newblock {\em J. Chem. Phys.}, 110:8919-8929, 1999.

\bibitem{nielsen2001statistical}
S.~Nielsen, R.~Kapral, and G.~Ciccotti.
\newblock Statistical mechanics of quantum-classical systems.
\newblock {\em J. Chem. Phys.}, 115:5805-5815, 2001.

\bibitem{Kapral:2001fk}
R.~Kapral.
\newblock Quantum-classical {D}ynamics in a {C}lassical {B}ath.
\newblock {\em J. Phys. Chem. A}, 105(12):2885--2889, 2001.

\bibitem{kisil2005quantum}
V.~V. Kisil.
\newblock A quantum-classical bracket from p-mechanics.
\newblock {\em Europhys. Lett.}, 72:873--879, 2005.

\bibitem{agostini2007we}
F.~Agostini, S.~Caprara, and G.~Ciccotti.
\newblock Do we have a consistent non-adiabatic quantum-classical mechanics?
\newblock {\em Europhys. Lett.}, 78:30001, 2007.

\bibitem{kisil2010comment}
V.~V. Kisil.
\newblock Comment on 'do we have a consistent non-adiabatic quantum-classical
  mechanics?" by {A}gostini {F}. et al.'.
\newblock {\em Europhys. Lett.}, 89:50005, 2010.

\bibitem{agostini2010reply}
F.~Agostini, S.~Caprara, and G.~Ciccotti.
\newblock Reply to the comment by {V.V. Kisil}.
\newblock {\em Europhys. Lett.}, 89:50006, 2010.

\bibitem{Schmitt1996JCP}
U.~Schmitt and J.~Brickmann.
\newblock Discrete time-reversible propagation scheme for mixed
  quantum-classical dynamics.
\newblock {\em Chem. Phys.}, 208:45-56, 1996.

\bibitem{kibble1979geometrization}
T.~W.~B. Kibble.
\newblock Geometrization of quantum mechanics.
\newblock {\em Com. Math. Phys.}, 65(2):189--201, 1979.

\bibitem{abbati1984pure}
M.~C. Abbati, R.~Cirelli, P.~Lanzavecchia, and A.~Mania.
\newblock Pure states of general quantum-mechanical systems as {K}{\"a}hler
  bundles.
\newblock {\em Nuo. Cim. B (1971-1996)}, 83(1):43--60, 1984.

\bibitem{cirelli1991quantum}
R.~Cirelli, A.~Mani{\`a}, and L.~Pizzocchero.
\newblock Quantum phase space formulation of schr{\"o}dinger mechanics.
\newblock {\em Int. J. Mod. Phys. A}, 6:2133--2146, 1991.

\bibitem{brody2001geometric}
D.~C. Brody and L.~P. Hughston.
\newblock Geometric quantum mechanics.
\newblock {\em J. of Geom. and Phys.}, 38(1):19--53, 2001.

\bibitem{Ashtekar:1998p906}
A.~Ashtekar and T.A. Schilling.
\newblock {\em On Einstein Path: essays in honor of Englebert Schucking, {\rm
  A. Harvey Eds.}}, chapter Geometrical Formulation of Quantum Mechanics, pages
  23--65.
\newblock Springer-Verlag, 1998.

\bibitem{Carinena:2006p7565}
J.~F. Cari{\~n}ena, J.~Clemente-Gallardo, and G.~Marmo.
\newblock {\em Proc. of the XV International Workshop on Geometry and Physics},
  volume~11, chapter Introduction to Quantum Mechanics and the
  Quantum-Classical transition, pages 3--45.
\newblock RSME, 2006.

\bibitem{Carinena:2007p813}
J.~F. Cari{\~n}ena, J.~Clemente-Gallardo, and G.~Marmo.
\newblock Geometrization of quantum mechanics.
\newblock {\em Theor. Math. Phys.}, 152(1):894--903, 2007.

\bibitem{clemente2008basics}
J.~Clemente-Gallardo and G.~Marmo.
\newblock Basics of quantum mechanics, geometrization and some applications to
  quantum information.
\newblock {\em Int. J. Geom. Meth. Mod. Phys.}, 5(6):989--1032, 2008.

\bibitem{balescu1997statistical}
R.~Balescu.
\newblock {\em Statistical dynamics: matter out of equilibrium}.
\newblock Imperial College Press, 1997.

\bibitem{balescu1975equilibrium}
R.~Balescu.
\newblock {\em Equilibrium and nonequilibrium statistical mechanics}.
\newblock Wiley-Interscience, 1975.

\bibitem{tolman}
R.~C. Tolman.
\newblock {\em The {P}rinciples of {S}tatistical {M}echanics}.
\newblock Oxford: Clarendon Press, 1938.

\bibitem{Andrade:2009p7108}
X.~Andrade, A.~Castro, D.~Zueco, J.~L. Alonso, P.~Echenique, F.~Falceto, and
  A.~Rubio.
\newblock Modified {E}hrenfest formalism for efficient large-scale ab initio
  molecular dynamics.
\newblock {\em J. Chem. Theor. Comp.}, 5:728--742, 2009.

\bibitem{Abraham:1978p1131}
R.~Abraham and J.~E. Marsden.
\newblock {\em Foundations of Mechanics}.
\newblock Reading, Massachusetts, 1978.

\bibitem{landsnab}
N.~P. Landsman.
\newblock {\em Mathematical topics between {C}lassical and {Q}uantum
  {M}echanics}.
\newblock Springer, 1998.

\bibitem{Oliveira2007ergodic}
C.~R. Oliveira and T.~Werlang.
\newblock Ergodic hypothesis in classical statistical mechanics.
\newblock {\em Revista Brasileira de Ensino de F\'isica}, 29(2):189--201, 2007.

\bibitem{Schwabl2002statistical}
F.~Schwabl.
\newblock {\em Statistical Mechanics}.
\newblock Springer, 2002.

\bibitem{Jaynes1957maxent1}
E.~T. Jaynes.
\newblock Information theory and {S}tatistical {M}echanics-{I}.
\newblock {\em Phys. Rev.}, 106(4):620--630, 1957.

\bibitem{Ellis1985largedev}
R.~S. Ellis.
\newblock {\em Entropy, large deviations and {S}tatistical {M}echanics}.
\newblock Spriger-Verlag, 1985.

\bibitem{Touchette2009largedev}
H.~Touchette.
\newblock The large deviation approach to statistical mechanics.
\newblock {\em Phys. Rep.}, 478:1--69, 2009.

\bibitem{Reinmann2008foundations}
P.~Reimann.
\newblock {Foundation of statistical mechanics under experimentally realistic
  conditions}.
\newblock {\em Phys. Rev. Lett.}, 101(19):190403, November 2008.

\bibitem{Reinmann2010njp}
P.~Reimann.
\newblock {Canonical thermalization}.
\newblock {\em New J. Phys.}, 12(5):055027, May 2010.

\bibitem{futurepaper}
J.~L. Alonso, A.~Castro, J.~Clemente-Gallardo, J.~C. Cuch\'i, P.~Echenique,
  F.~Falceto, and D.~Zueco.
\newblock In preparation.

\end{thebibliography}

\end{document}